\documentclass[
twocolumn,
aps,
prl,
reprint,
superscriptaddress,
amsmath,amssymb,
]{revtex4-1}
\usepackage[pdftex]{graphicx, color}
\usepackage[pdftex,
            colorlinks=true,
            citecolor=blue,
            linkcolor=blue]{hyperref}
\graphicspath{{./fig/}{./fig_phase_diagram_t20/}{./fig_phase_diagram_t202/}}

\usepackage{braket}
\usepackage{times,multirow,amsfonts,bm,xspace,pifont}
\usepackage{soul}
\usepackage[normalem]{ulem}

\begin{document}
\newcommand{\rr}{{\bm r}}
\newcommand{\q}{{\bm q}}
\renewcommand{\k}{{\bm k}}
\newcommand*\wien    {\textsc{wien}2k\xspace}
\newcommand*\textred[1]{\textcolor{red}{#1}}
\newcommand*\textblue[1]{\textcolor{blue}{#1}}
\newcommand{\ki}[1]{{\color{red}\st{#1}}}
\newcommand{\sgn}{\mathrm{sgn}\,}
\newcommand{\tr}{\mathrm{tr}\,}
\newcommand{\Tr}{\mathrm{Tr}\,}
\newcommand{\GL}{{\mathrm{GL}}}
\newcommand{\talpha}{{\tilde{\alpha}}}
\newcommand{\tbeta}{{\tilde{\beta}}}
\newcommand{\mathN}{{\mathcal{N}}}
\newcommand{\mathQ}{{\mathcal{Q}}}
\newcommand{\bv}{{\bar{v}}}
\newcommand{\bj}{{\bar{j}}}

\newcommand{\YY}[1]{\textcolor{magenta}{#1}}
\newcommand{\AD}[1]{\textcolor{blue}{#1}}
\newcommand*{\ADS}[1]{\textcolor{blue}{\sout{#1}}}
\newcommand*\YYS[1]{\textcolor{magenta}{\sout{#1}}}
\newcommand{\reply}[1]{{#1}}
\newcommand{\replyS}[1]{\textcolor{red}{\sout{#1}}}

% Title of paper
\title{Intrinsic Superconducting Diode Effect}

\author{Akito Daido}
\affiliation{Department of Physics, Graduate School of Science, Kyoto University, Kyoto 606-8502, Japan}
\email[]{daido@scphys.kyoto-u.ac.jp}
\author{Yuhei Ikeda}
\affiliation{Department of Physics, Graduate School of Science, Kyoto University, Kyoto 606-8502, Japan}
\author{Youichi Yanase}
\affiliation{Department of Physics, Graduate School of Science, Kyoto University, Kyoto 606-8502, Japan}
\affiliation{%
  Institute for Molecular Science, Okazaki 444-8585, Japan
}%
\date{\today}

\begin{abstract}
Stimulated by the recent experiment [F. Ando \textit{et al}., Nature \textbf{584}, 373 (2020)], we propose an intrinsic mechanism to cause the superconducting diode effect (SDE).
SDE refers to the nonreciprocity of the critical current for the metal-superconductor transition.
Among various mechanisms for the critical current, the depairing current is known to be intrinsic to each material and has recently been observed in several superconducting systems.
We clarify the temperature scaling of the nonreciprocal depairing current near the critical temperature and point out its significant enhancement at low temperatures.
It is also found that the nonreciprocal critical current shows sign reversals upon increasing the magnetic field.
These behaviors are understood by the nonreciprocity of the Landau critical momentum and the \reply{change in the nature} of the helical superconductivity.
The intrinsic SDE unveils the rich phase diagram and functionalities of noncentrosymmetric superconductors.
\end{abstract}

\maketitle

\textit{Introduction}. --- 
Rectification by the semiconductor diode is one of the central building blocks of electronic devices.
Apart from the nonreciprocity induced by asymmetric junctions, it has been revealed that nonreciprocal transport can be obtained as a bulk property of materials~\cite{Tokura2018-nb,Ideue2021-es}.
Magnetochiral anisotropy (MCA)~\cite{Rikken2001-il,Krstic2002-vo,Pop2014-wn,Rikken2005-ew,Ideue2017-vp,Wakatsuki2018-ll,Hoshino2018-sa,Wakatsuki2017-dp,Qin2017-vd,Yasuda2019-jw,Itahashi2020-ef}
is an example, described by the equation
%\begin{equation}
$R(j)=R_0(1+\gamma j\reply{h}).$
%\end{equation}
Here $R$, $j$, and $\reply{h}$ are the resistance, electric current, and the magnetic field, respectively.
The coefficient $\gamma$ gives rise to different resistance for rightward and leftward electric currents and can be finite in noncentrosymmetric materials.
MCA has been observed in (semi)conductors~\cite{Rikken2005-ew,Ideue2017-vp, Krstic2002-vo,Pop2014-wn} as well as in superconductors~\cite{Wakatsuki2017-dp,Qin2017-vd,Yasuda2019-jw,Itahashi2020-ef},
and allows us to access various aspects of noncentrosymmetric materials: from spin-orbit splitting in the band structure~\cite{Ideue2017-vp} to the spin-singlet and -triplet mixing of Cooper pairs~\cite{Wakatsuki2017-dp,Wakatsuki2018-ll,Hoshino2018-sa}
%,Itahashi2020-ef}.

\begin{figure}
    \centering
    \includegraphics[width=0.425\textwidth]{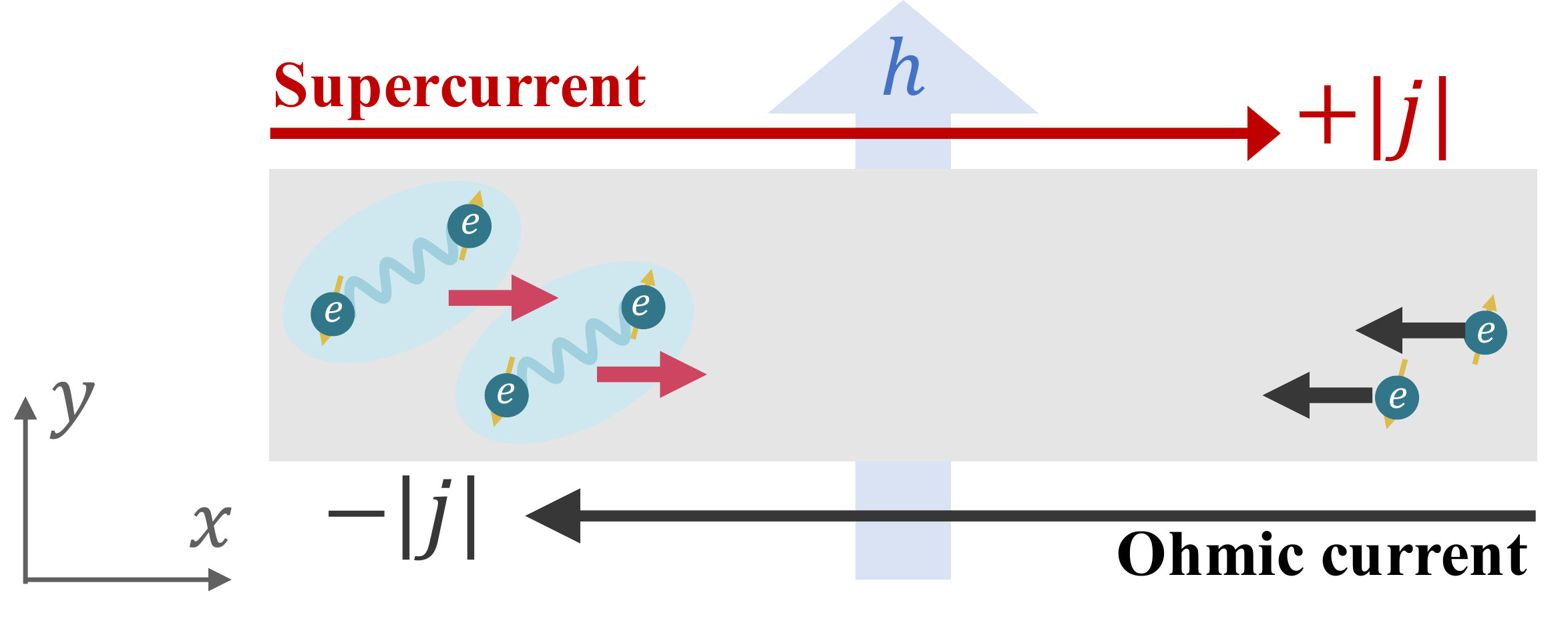}
    \caption{Schematic figure for the SDE. The system has zero and finite resistance for the rightward and leftward current, respectively, and \textit{vice versa} when the magnetic field $h$ is reversed.}
    \label{fig:schematic}
\end{figure}

MCA is the inequivalence of $R(j)$ and $R(-j)$, where both $R(\pm j)$ usually take finite values.
On the other hand, such a drastic situation is possible in superconductors that either one of $R(\pm j)$ vanishes while the other remains finite [Fig.~\ref{fig:schematic}].
Such a \textit{superconducting diode effect} (SDE) has recently been observed in the Nb/V/Ta superlattice without an inversion center and is controlled by the applied inplane magnetic field~\cite{Ando2020-om}.
{This is the first report of the SDE in a bulk material, while similar effects have been recognized in engineered systems~\cite{Reynoso2008-or,Zazunov2009-we,Margaris2010-ax,Yokoyama2014-fw,Silaev2014-bp,Campagnano2015-jl,Dolcini2015-cu,Chen2018-gz,Minutillo2018-dz,Pal2019-ed,Kopasov2021-xz}
and followed by recent SDE experiments~\cite{Baumgartner2021-lg,Lyu2021-sm}.}
%\ADS{Subsequent experiments have also \AD{reported} \ADS{revealed} SDE in a Josephson junction~\cite{Baumgartner2021-aq} and a nanostructured system~\cite{Lyu2021-sm}.}
SDE is a promising building block of the dissipationless electric circuits, 
and is a fascinating phenomenon manifesting the interplay of the inversion breaking and superconductivity.
One of the remaining issues is to identify suitable materials providing the best performance;
however, the mechanisms to cause SDE in a bulk material~\cite{Ando2020-om} have not been clarified,
while the SDE in artificial devices~\cite{Lyu2021-sm,Baumgartner2021-lg} has been well simulated by Bogoliubov-de Gennes (BdG)~\cite{Baumgartner2021-lg} and time-dependent Ginzburg-Landau (GL) theories~\cite{Lyu2021-sm}.

SDE is the nonreciprocity of the critical current for the resistive transition.
In usual situations, in particular, under out-of-plane magnetic fields, the resistive transition is caused by the vortex motion.
The details of the vortex motion depend on the device setup such as impurity concentrations~\cite{Blatter1994-ug}, and in turn, has an advantage of tunability by the nanostructure engineering~\cite{Villegas2003-nc,Lyu2021-sm}.
Apart from the extrinsic mechanisms to cause resistivity,
the \textit{depairing current} is known as the critical current unique to each superconducting material. Here, the metal-superconductor transition is literally caused by the dissociation of the flowing Cooper pairs~\cite{Tinkham2004-dh,Dew-Hughes2001-ka}.
The depairing current always gives the upper limit of the critical current and is an important material parameter characterizing superconductors~\cite{Blatter1994-ug}.
The depairing limit generally requires a huge current density, but is within the scope of experimental techniques. Indeed, the depairing limit has recently been achieved in the %\ADS{nanobridge} 
{microbridge} superconducting devices of YBa$_2$Cu$_3$O$_{7-\delta}$~\cite{Nawaz2013-wp}, Ba$_{0.5}$K$_{0.5}$Fe$_2$As$_2$~\cite{Li2013-qq}, and Fe$_{1+y}$Te$_{1-x}$Se$_x$~\cite{Sun2020-ep}.

In this Letter, as a first step of the theoretical research on SDE of bulk materials, we propose the intrinsic mechanism of SDE by studying the nonreciprocity in the depairing current.
The results can be tested with the %\ADS{future} \ADS{nanobridge} 
{microbridge} experiments
%\ADS{, while might explain some qualitative aspects of Refs.}%~\cite{Ando2020-om,Ono_private}.
{and establish the foundation of the future study on the bulk SDE.}
Furthermore, it is revealed that the intrinsic SDE is closely related to the Flude-Ferrell-Larkin-Ovchinnikov state~\cite{Larkin1964-en,Fulde1964-qq}.
While the Larkin-Ovchinnikov (LO) state with the spatially inhomogeneous pair potential $\Delta(x)= \Delta \cos qx$ has been discussed for FeSe~\cite{Kasahara2014-zz,Kasahara2020-xw}, CeCoIn$_5$~\cite{Matsuda2007-em} and organic superconductors~\cite{Wosnitza2018-qh},
the Flude-Fellel (FF) type order parameter $\Delta(x)=\Delta e^{iqx}$ %(and LO-type one as well) 
is known to ubiquitously appear in noncentrosymmetric superconductors and is particularly called the \textit{helical superconductivity}~\cite{Bauer2012-xi,Smidman2017-hb,Agterberg2003-jn,Barzykin2002-eh,Dimitrova2003-mo,Kaur2005-jf,Agterberg2007-vl,Dimitrova2007-hp,Samokhin2008-nv,Yanase2008-yb,Bauer2012-xi,Michaeli2012-gl,Sekihara2013-dm,Houzet2015-iy}.
Implications of the helical superconductivity have been obtained in thin films of Pb~\cite{Sekihara2013-dm} and doped SrTiO$_3$~\cite{Schumann2020-mw}, and a heavy-fermion superlattice~\cite{Naritsuka2017-or,Naritsuka2021-ym}.
We show that the intrinsic SDE works as a probe to study the phase diagram of helical superconductivity.
{Relation to the recent experiments~\cite{Ando2020-om,Ono_private} is also discussed.}

%%%%%%%%%%%%%%%%%%%%%%%%%%%%%%%%%%%%%%%%%%%%%%%%%%%%
\textit{Model.} --- 
We consider the critical current in two-dimensional (2D) superconductors with a polar axis due to the substrate and/or the crystal structure.
The magnetic field is applied along the $y$ direction, which makes the critical current nonreciprocal in the $x$ direction [Fig.~\ref{fig:schematic}].
The system is modeled by the Rashba-Zeeman Hamiltonian with the attractive Hubbard interaction,
\begin{align}
\hat{H}&=\sum_{\bm k \sigma\sigma'}\bigl[\xi (\bm k)\delta_{\sigma\sigma'}+\bm{g}(\bm k) \cdot \bm{\sigma}_{\sigma \sigma'}
- h(\sigma_y)_{\sigma \sigma'}\bigr]
c_{\bm k \sigma}^\dagger c_{\bm k \sigma'} 
\nonumber \\
&\qquad\quad-\frac{U}{V}\sum_{\bm{k}_1+\bm{k}_2+\bm{k}_3+\bm{k}_4=\bm{0}}c^\dagger_{\bm{k}_1\uparrow}c^\dagger_{\bm{k}_2\downarrow}c_{\bm{k}_3\downarrow}c_{\bm{k}_4\uparrow}.
\label{eq:model}
\end{align}
Here, $\xi(\bm{k})=-2t_1(\cos k_x+\cos k_y)+4t_2\cos k_x\cos k_y-\mu$ and $\bm{g}(\bm{k})=\alpha_g(-\sin k_y,\,\sin k_x,\,0)$ represent the hopping energy and the Rashba spin-orbit coupling, respectively.
The magnetic field in the $y$ direction is introduced by the Zeeman term $h\equiv\mu_BH_y$.
The parameters are given by $(t_1,t_2,\mu,\alpha_g,U)=(1,0,-1,0.3,1.5)$
unless mentioned otherwise.
The next-nearest-neighbor hopping $t_2$ is introduced for the latter use.
The energy dispersion of the noninteracting part is given by
% \begin{align}
%     \xi_\chi^h(\bm{k})&=\xi(\bm{k})+\chi\sqrt{g_x(\bm{k})^2+(g_y(\bm{k})-h)^2}\notag\\
%     &=\xi^0_\chi(\bm{k}-\bm{q}_\chi(\bm{k})/2)+O(h^2).
% \end{align}
$\xi_\chi^h(\bm{k})=\xi(\bm{k})+\chi|\bm{g}(\bm{k})-h\hat{y}|\simeq\xi^0_\chi(\bm{k}-\bm{q}_\chi(\bm{k})/2)$.
Here, each band is labeled by the helicity $\chi=\pm$, and the momentum shift under $h$ is given by 
$\bm{q}_\chi(\bm{k})/2=\chi g_y(\bm{k})h\bm{v}_\chi(\bm{k})/|\bm{v}_\chi(\bm{k})|^2$
with $\bm{v}_\chi(\bm{k})\equiv\nabla\xi^0_\chi(\bm{k})$.
The momentum shift is estimated by its Fermi-surface (FS) average, $q_\chi\equiv\hat{x}\cdot\braket{\bm{q}_\chi(\bm{k})}_{\mathrm{FS}}\sim2\chi h/\braket{|\bm{v}_\chi(\bm{k})|}_{\mathrm{FS}}$.

We solve the model~\eqref{eq:model} within the mean-field approximation.
The attractive Hubbard interaction is approximated by
\begin{align}
&{\frac{1}{2}}\sum_{\bm{k}\sigma\sigma'}\Delta (i\sigma_y)_{\sigma\sigma'} c^\dagger_{\bm{k}+\bm{q}\sigma}c^\dagger_{-\bm{k}\sigma'} +\mathrm{H.c.}+\Delta^2/2U.
\label{eq:MF_main}
\end{align}
The $s$-wave pair potential $\Delta$ is considered with a center-of-mass momentum $\bm{q}=q\hat{x}$ to describe the current-flowing state. 
For a given $q$, the value of $\Delta=\Delta(q)$ is determined self-consistently by the gap equation with the temperature $T$.
To describe the superconducting transitions and the supercurrent, it is convenient to introduce the condensation energy $F(q)$ for each $q$, that is, the difference of the free energy per unit area in the normal and superconducting states.
The sheet current density is obtained by $j(q)=2\partial_q F(q)$, which coincides with the expectation value of the current operator~\cite{Supplement}.

When an electric current $j_{\mathrm{ex}}$ is applied, the superconducting state with $q$ satisfying $j(q)=j_{\mathrm{ex}}$ should be realized.
However, no superconducting state can sustain $j_{\mathrm{ex}}$ when $j_{\mathrm{ex}}<j_{c-}$ or $j_{\mathrm{ex}}>j_{c+}$, with
$j_{c+}\equiv\max_qj(q)$ and $j_{c-}\equiv\min_qj(q).$
Thus, the depairing current in the positive and negative directions is given by the maximum $j_{c+}$ and minimum $j_{c-}$ of $j(q)$, respectively.
In particular, the nonreciprocal component is given by
\begin{equation}
    \Delta j_c\equiv j_{c+}+j_{c-}= j_{c+}-|j_{c-}|.
\end{equation}
The SDE is identified with a finite $\Delta j_c$ of the system.
We also define the averaged critical current $\bar{j}_c\equiv (j_{c+}-j_{c-})/2$, by which the strength of the nonreciprocal nature can be expressed as $r\equiv\Delta j_c/\bar{j}_c$.

\textit{GL analysis.} --- 
First, we discuss the SDE by the GL theory.
The GL free energy %,
%\begin{equation}
    $f(\Delta,q)=\alpha(q)\Delta^2+\frac{\beta(q)}{2}\Delta^4$
%\end{equation}
gives a good approximation of $F(q)$ near the transition temperature $T_c$ when the optimized order parameter $\Delta=\Delta(q)$ is substituted.
The GL coefficients are assumed to have the following form:
%\begin{gather}
    $\alpha(q)=\alpha_0+\alpha_1 q+{\frac{1}{2}}\alpha_2 q^2+{\frac{1}{6}}\alpha_3q^3$, %\ 
    and $\beta(q)=\beta_0+\beta_1q$,
%    \label{eq:standard_GL}
%\end{gather}
which is valid for the description up to $O(T_c-T)^{5/2}$.
When the higher-order gradient terms $\alpha_3,\beta_1$ are neglected,
the broken inversion and time-reversal symmetries are encoded solely into $\alpha_1\neq0$,
which shifts the minimum of $f(q)=f(\Delta(q),q)$ from $q=0$ to $q_0=-\alpha_1/2\alpha_2$.
Thus, the superconducting state with a finite $q_0$, namely the helical superconductivity is realized~\cite{Agterberg2003-jn,Smidman2017-hb}.
The helical superconducting state with $q=q_0$ does not carry a supercurrent~\cite{Agterberg2003-jn,Dimitrova2003-mo,Dimitrova2007-hp}, 
%\begin{equation}
    $j(q_0)=2\partial_{q_0}f(q_0)=0,$
%\end{equation}
as the most stable state generally should be.

It is convenient to rewrite the GL coefficients
%Eq.~\eqref{eq:standard_GL} 
as %the following form:
%\begin{gather}
    $\alpha(q)=\talpha_0+\frac{\tilde{\alpha}_2}{2}(q-\tilde{q}_0)^2+\frac{{\alpha}_3}{6}(q-\tilde{q}_0)^3$ %,\notag\\
    and $\beta(q)=\tbeta_0+\beta_1(q-\tilde{q}_0)$,
%\end{gather}
where the linear term in $\alpha(q)$ is erased.
Clearly, $f(\Delta,q+\tilde{q}_0)$ for $\alpha_3=\beta_1=0$ is equivalent to the GL free energy of a centrosymmetric superconductor, leading to a reciprocal critical current~\cite{Smidman2017-hb}.
Thus, the SDE is caused by the higher-order terms, $\alpha_3$ and $\beta_1$, \begin{align}
    \Delta j_c&=\left(\frac{16}{27\tbeta_0\talpha_2}\alpha_3-\frac{8}{9\tbeta_0^2}\beta_1\right)\talpha_0^2,\label{eq:Djc_GL}
\end{align}
up to first order in $\alpha_3$ and $\beta_1$~\cite{Supplement}.
Note that $\Delta j_c\propto (T_c-T)^2$ in contrast to the averaged critical current $\bar{j}_c\propto (T_c-T)^{3/2}$~\cite{Tinkham2004-dh,Supplement}, since $\talpha_0\propto T-T_c$.
Thus, a small but finite $\Delta j_c$ is predicted by the GL theory, while a larger $\Delta j_c$ is expected at low temperatures.
The result obtained here is valid for general noncentrosymmetric superconductors without orbital depairing effect, e.g., superconducting thin films under inplane magnetic fields.

\begin{figure}[t]
    \centering
        \includegraphics[width=0.475\textwidth]{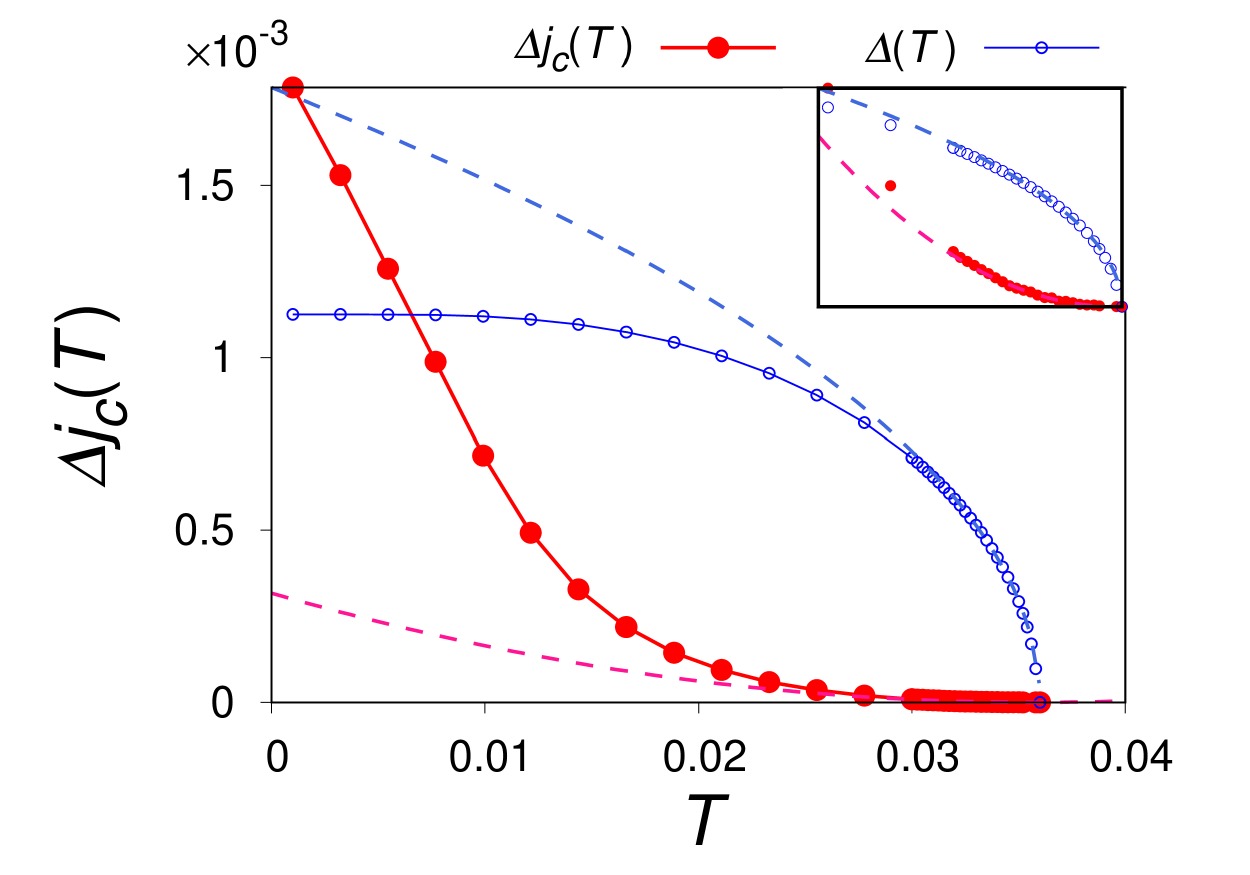}
    \caption{
    The temperature dependence of $\Delta j_c$ at $h=0.03$.
    The red closed circles indicate $\Delta j_c(T)$, while the open blue circles indicate $\Delta(T)$ (a.u.).
    The dashed lines show the fitting curve of $\Delta j_c(T)$ and $\Delta(T)$ near $T_c$ with $(T_c-T)^2$ and $\sqrt{T_c-T}$, respectively.
    The inset shows the enlarged figure near $T_c\simeq 0.036$.}
    \label{fig:jc_vs_T}
\end{figure}

\textit{{Critical current under low fields.} ---}
Equipped with the insight of the GL theory, we discuss the temperature dependence of $\Delta j_c$ based on the model~\eqref{eq:model}. 
The result is shown in Fig.~\ref{fig:jc_vs_T} for $h=0.03$.
% \ADS{(-12)The closed red and open blue circles indicate $\Delta j_c(T)$ and $\Delta(T)$, respectively.}
As shown in the inset, the temperature scaling $\Delta j_c\propto(T_c-T)^2$ is confirmed near the transition temperature $T_c$.
% \ADS{(-9)As expected, $\Delta j_c(T)$ deviates from the GL prediction $\Delta j_c(T) \propto (T_c-T)^2$}
{The scaling law becomes inaccurate} as $T_c-T$ gets large, where $\Delta(T)$ also deviates from $\Delta(T) \propto \sqrt{T_c-T}$.
%\ADS{(-2)In particular}  
{Importantly}, $\Delta j_c(T)$ is strongly enhanced at low temperatures.

\begin{figure}[t]
    \centering
\includegraphics[width=0.5\textwidth]{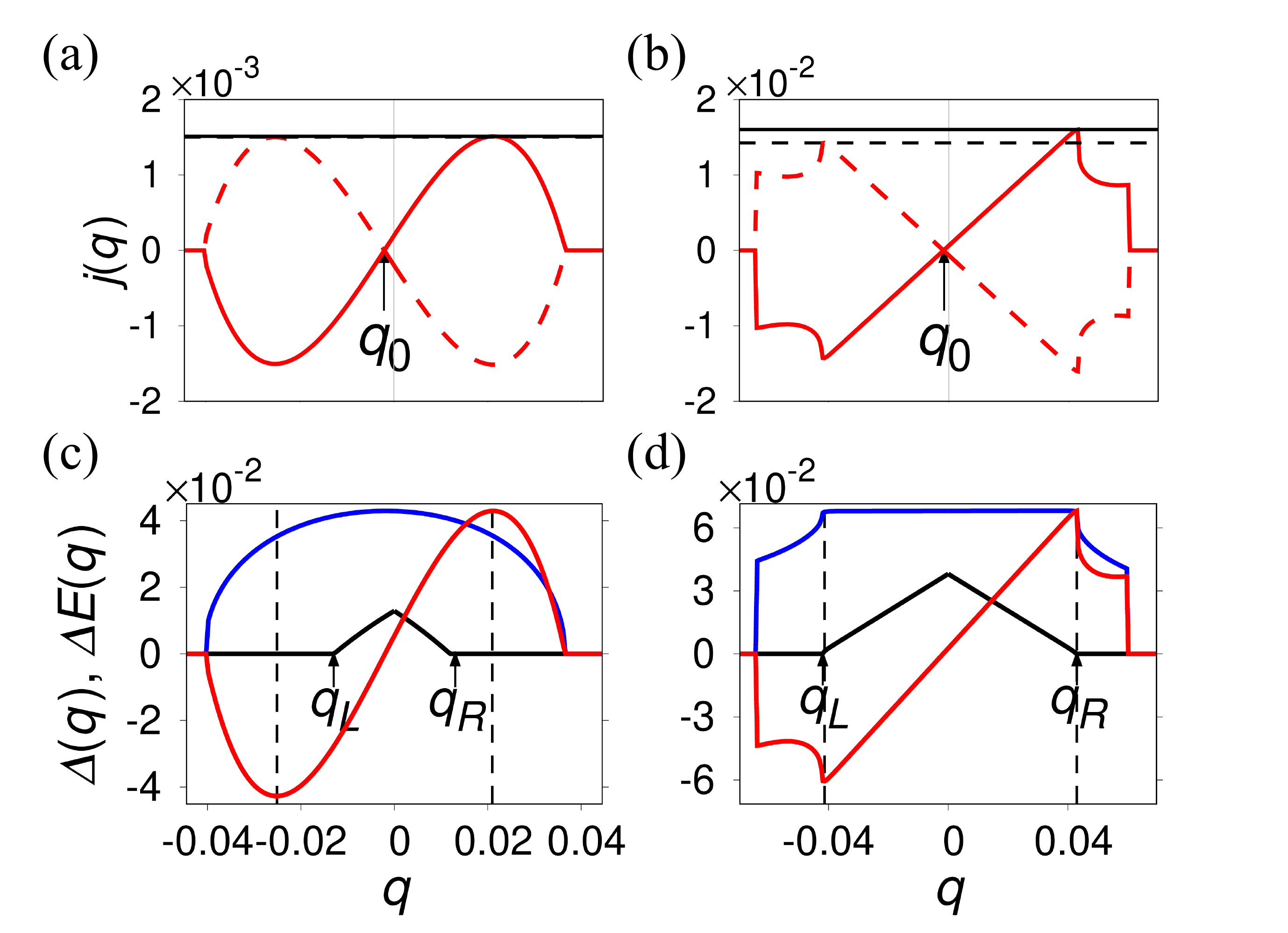}
    \caption{
    (a), (b): The $q$ dependence of the supercurrent at $h=0.03$ and (a) $T=0.03$ and (b) $T=0.001$.
    In addition to $j(q)$ (red lines), $-j(q)$ (dashed red lines) is shown. The black (dashed) horizontal lines indicate $j_{c+}$ ($-j_{c-}$).
    %The inset in the panel (a) shows the enlarged figure of $j(q),-j(q)\simeq j_{c+}$. 
    The position of $q_0$ is indicated by arrows.
    (c), (d): The order parameter $\Delta(q)$ (blue lines) and the excitation gap $\Delta E(q)$ (black lines) are shown together with $j(q)$ (a.u.).
    The parameters for the panels (c) and (d) are the same as the panel (a) and (b), respectively.
    %$\Delta(q)$ and $\Delta E(q)$ are shown by open blue circles and closed gray squares, respectively.
    The vertical dashed black lines indicate the momentum $q$ where $j(q)=j_{c\pm}$.
    Landau critical momentum $q_R$ and $q_L$ are indicated by arrows.
    }
    \label{fig:NR_Landau_all}
\end{figure}

To clarify the origin of the SDE, we show $j(q)$ by red lines in Figs.~\ref{fig:NR_Landau_all}. 
In Fig.~\ref{fig:NR_Landau_all}~(a) for $T=0.03 \simeq T_{\rm c}$, $j(q)$ is a smooth curve and its tiny asymmetry gives rise to $\Delta j_c$, as is illustrated by the difference of the solid and dashed horizontal lines (indicating $j_{c+}$ and $-j_{c-}$).
This is consistent with the GL picture where $\Delta j_c$ is caused by the asymmetry factors $\alpha_3,\beta_1\neq0$.
Two curves, $j(q)$ and $-j(q)$, cross at $q_0<0$, indicating the helical superconductivity.
In Fig.~\ref{fig:NR_Landau_all}~(c), $\Delta(q)$ and the minimum excitation energy $\Delta E(q)$ are shown in addition to $j(q)$, by the blue and black lines, respectively.
The superconducting state remains stable even after the spectrum becomes gapless, and reaches the maximum and minimum of $j(q)$ in the gapless region.

As shown in Fig.~\ref{fig:NR_Landau_all}~(b), 
the dispersion of $j(q)$ at $T=0.001 \ll T_{\rm c}$ is significantly different from that at $T=0.03$, and
a large $\Delta j_c$ is realized.
The maximum and minimum of $j(q)$ are achieved at the ends of the region where $j(q)$ 
is almost linear in $q$.
These momenta approximately coincide with the Landau critical momenta, $q_R>0$ and $q_L<0$, i.e. the first $q$'s satisfying $\Delta E(q)=0$, as is clear from Fig.~\ref{fig:NR_Landau_all}~(d).
Actually, the depairing effect takes place after $q>q_R$ or $q<q_L$: The excited quasiparticles reduce $|j(q)|$ and $\Delta(q)$ and finally cause a first-order phase transition into the normal state.
From these observations, we obtain the formula
% \begin{gather}
%     j_{c+}=n^s_{xx}(q_R-q_0)/2,\quad j_{c-}=n^s_{xx}(q_L-q_0)/2,\notag\\
%     \Delta j_c= n^s_{xx}(q_R+q_L-2q_0)/2\label{eq:Landau_Djc},
% \end{gather}
\begin{equation}
    \Delta j_c= n^s_{xx}(q_R+q_L-2q_0)/2\label{eq:Landau_Djc},
\end{equation}
by using, e.g., $j_{c+}=n^s_{xx}(q_R-q_0)/2$ with
the superfluid {weight} $n_{xx}^s=2\partial_{q_0}j(q_0)$.
Thus, \textit{the nonreciprocal Landau critical momentum} $q_R+q_L$ \textit{measured from} $2q_0$ gives rise to the SDE at extremely low temperatures.
As $T$ gets larger, the maximum and minimum of $j(q)$ deviate from $j(q_R)$ and $j(q_L)$, and Eq.~\eqref{eq:Landau_Djc} becomes no longer valid.
The mechanism of the SDE at low temperatures is not captured by the GL theory.

\begin{figure}[t]
    \centering
\includegraphics[width=0.5\textwidth]{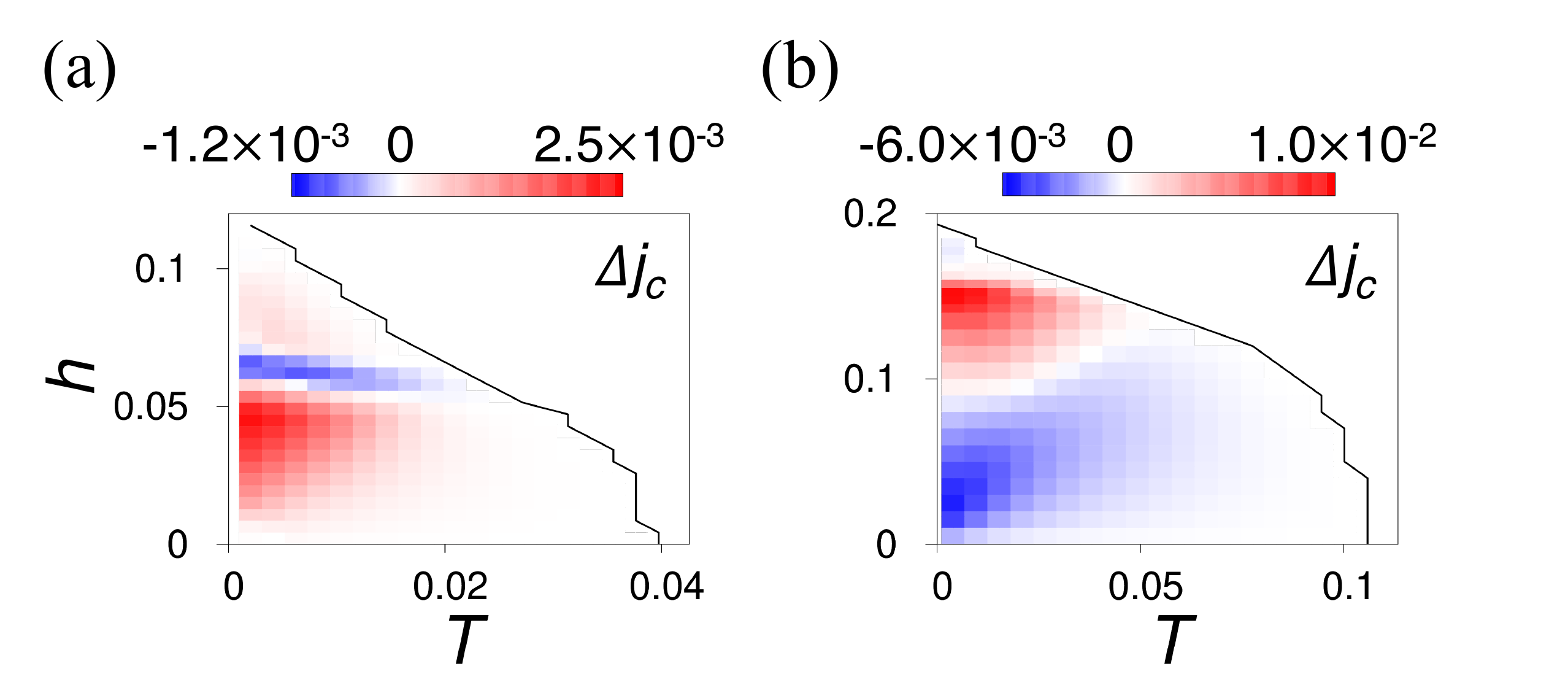}
    \caption{
    The magnetic-field and temperature dependence of the nonreciprocal component of the critical current $\Delta j_c(h,T)$  for (a) $t_2=0$ and (b) $t_2=0.2$.
%    The vertical and horizontal axes are $h$ and $T$, respectively.
The red and blue colors indicate positive and negative values of $\Delta j_c$, respectively.
    The transition temperature $T_c(h)$ determined with the $T$ mesh (a) $0.045/21$ and (b) $0.12/21$ is shown with the black line for the guide of the eye.
    }
    \label{fig:Djc_HT}
\end{figure}

\textit{{Phase diagram.} ---}
In Figs.~\ref{fig:Djc_HT}~(a) and (b), we show the temperature and magnetic-field dependence of the nonreciprocal component $\Delta j_c$ for $t_2=0$ and $t_2=0.2$.
Let us focus on the low-field region, where positive and negative values of $\Delta j_c$ are widely obtained for Fig.~\ref{fig:Djc_HT}~(a) and (b), respectively.
The sign reversal of $\Delta j_c$ by $t_2$ can be understood based on Eq.~\eqref{eq:Landau_Djc}.
Indeed, we show in the Supplemental Material~\cite{Supplement} that $q_R+q_L-2q_0$ causes a sign reversal as $t_2$ increases, leading to that of $\Delta j_c$ as well.
It is also shown that for large values of $t_2$, $q_R+q_L-2q_0$ is dominated by the nonreciprocal Landau critical momentum $q_R+q_L$, while it is dominated by $-2q_0$ for small values of $t_2$.
A relatively large SDE for $t_2\sim 0.2$ is explained by large values of $q_R+q_L-2q_0$ as a result of the anisotropy~\cite{Supplement}.
% \YYS{as result of the anisotropy of the system: In particular, the difference of the maximum Fermi velocity between two helicity bands plays an important role~\cite{Supplement}.(25)}
The pronounced aspect of Fig.~\ref{fig:Djc_HT} is the sign reversals prevailing under moderate and high magnetic fields. This point will be discussed in the following.

\textit{Critical current under high fields.} ---
\begin{figure}
    \centering
    \includegraphics[width=0.5\textwidth]{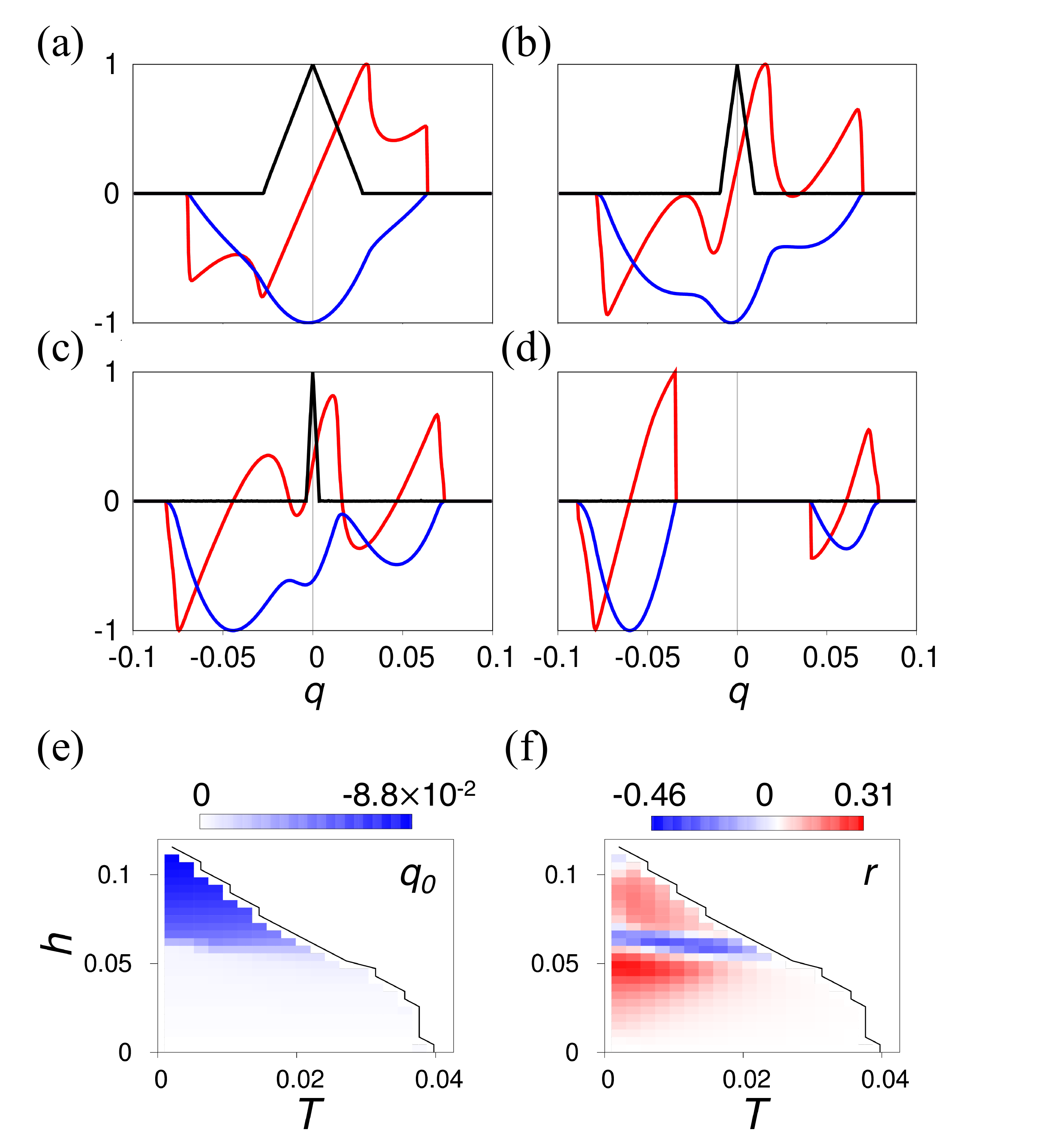}
    \caption{
    (a)-(d) $j(q)$ (red lines), $F(q)$ (blue lines), and $\Delta E(q)$ (black lines) normalized to $[-1,1]$  at $T=0.001$ for (a) $h=0.043$, (b) $h=0.058$, (c) $h=0.063$ and (d) $h=0.075$.
    (e),(f) $q_0$ and $r$ for various values of $h$ and $T$.
    }
    \label{fig:three_peaks}
\end{figure}
To see the origin of the high-field behavior, we show the condensation energy $F(q)$ at $T=0.001$ for various values of $h$ in Fig.~\ref{fig:three_peaks}.
To be specific, the case of Fig.~\ref{fig:Djc_HT} (a) is considered.
The condensation energy $F(q)$ shown by the blue line has a single-{well} structure under {low magnetic fields} [panel (a)].
% \ADS{(-32)The structure near $|q|\sim0.05$ is developed under higher magnetic fields, and there appear two side peaks under $h=0.058$ [panel (b)], and subsequently the left peak becomes most stable under $h=0.063$ [panel (c)].}
{The structure near $|q|\sim0.05$ is developed under higher magnetic fields [panel (b)], to form two local minima [panel (c)], where the left one becomes most stable.}
These side {wells} are the precursor of the high-field helical superconducting states [panel (d)], where the central {minimum} finally disappears.
Such a \reply{change} is most evident in $q_0(T,h)$ shown in Fig.~\ref{fig:three_peaks}~(e).
Under low fields, $q_0$ is determined by the balance of two Fermi surfaces shifted in the opposite directions, resulting in $|q_0|\lesssim 10^{-2}$; on the other hand, under high fields, $q_0\sim 10^{-1}$ almost coincides with $q_\chi$ of the Fermi surface with a larger density of states~\cite{Dimitrova2003-mo,
Agterberg2007-vl,Smidman2017-hb,Yanase2008-yb}.
This determines the \reply{``crossover line"}
\footnote{\reply{This terminology is named after the crossover generally seen in noncentrosymmetric superconductors, while the line changes to the first-order transition at lower temperature $T\lesssim0.12$ in this model, as in Figs.5(b)-5(d).}}
of helical superconductivity visible at $h\sim0.06$ in Fig.~\ref{fig:three_peaks}~(e).
% \reply{This terminology is named after the crossover generally seen in noncentrosymmetric superconductors, while the line changes to the first-order transition at lower temperature $T\lesssim@@@$ in this model, as in Fig.5.}

The evolution of $j(q)$ by $h$ follows that of $F(q)$.
Overall,  $j(q)$ consists of several almost-straight lines and their interpolation, since
$F(q)$ is approximated by the square function of $q$ around each local minimum.
Comparing Figs.~\ref{fig:three_peaks} (a) and (b), the $q$ point achieving $j_{c-}$ is changed from $q_L$ to a critical momentum of the left {well}, which we name $q_{-,L}$.
In the panel (c), $q_{-,L}$ remains to give $j_{c-}$, whose value is significantly enhanced owing to the development of the left {minimum}.
This causes the sign reversal of $\Delta j_c$.
In the panel (d), $\Delta j_c$ is determined by the tiny asymmetry of the left {well}.
It should be noticed that the ratio $r=\Delta j_c/\bar{j}_c$ shown in Fig.~\ref{fig:three_peaks}~(f) is quite large around the crossover line, takes values up to $|r|\lesssim 0.4$, as is understood from Figs.~\ref{fig:three_peaks}~(a)-(c).
Thus, the sign reversals and huge values of $r=\Delta j_c/\bar{j}_c$ under magnetic fields are caused by \reply{the change in} the helical superconducting states.
According to Figs.~\ref{fig:three_peaks}~(e) and (f), 
\reply{the sign reversal also occurs near $T_c(h)$ by the crossover.}

Figure~\ref{fig:Djc_HT}~(b) can be understood similarly.
In this case, the crossover line is identified to be $h\sim 0.17$~\cite{Supplement}, where $\Delta j_c$ changes its sign.
The high-field helical superconducting states span only a small fraction in the phase diagram.
The difference from Fig.~\ref{fig:Djc_HT}~(a) is another sign reversal at $h\sim0.09$.
In this region, $\Delta j_c$ is determined by the nonreciprocal Landau critical momentum, and $q_R+q_L-2q_0$ turns out to change its sign by increasing $h$.
This is because $q_R+q_L(<0)$ shows nonmonotonic behavior, while $-2q_0(>0)$ grows linearly and finally becomes dominant~\cite{Supplement}.
The sign reversal survives at higher temperatures and reaches the transition temperature.

\textit{Discussion} ---
We have revealed the sign reversals of the SDE, which is closely connected with the \reply{change in} the helical superconducting states.
Thus, the intrinsic SDE is a promising bulk probe directly unveiling the crossover line.
% \YYS{It has been proposed that helical superconductivity can be detected with Josephson-junction experiments~\cite{Kaur2005-jf}, and with
% the scanning-tunneling microscopy by observing the gapless spectrum in the high-field phase~\cite{Smidman2017-hb}.
% These probes are complimentary to study the helical superconductivity.($+17$ if recovered)}
This probe is complementary to the junction~\cite{Kaur2005-jf} and spectroscopy~\cite{Smidman2017-hb} experiments proposed to detect the helical superconductivity. 

In the end, we briefly discuss the connection with the experimental results of SDE~\cite{Ando2020-om}.
%\ADS{$\Delta j_c$ tends to be larger as lowering the temperature, which is also the case for the intrinsic SDE.}
The sign reversals of $\Delta j_c$ by increasing the magnetic field at low temperatures have recently been observed~\cite{Ono_private}, which might be explained by our results for the intrinsic SDE.
An ``inverse effect," the nonreciprocity of the critical magnetic field under applied electric current, has also been reported~\cite{miyasaka2021-ly}, implying the nonreciprocity as a bulk property of the superconductor.
Thus, the SDE with sign reversals implies the crossover in the superconducting state of the Nb/V/Ta superlattice.
On the other hand, 
%\AD{the observed $\Delta j_c$ has weak temperature dependence within $0.6\, T_c\lesssim T$, and experiments under lower temperature are awaited~\cite{Ono_private}.}
% \ADS{(-29)the magnetic-field dependence of $\Delta j_c$ near the transition temperature is at variance with the intrinsic SDE: $\Delta j_c$ jumps from $H=-0$ to $H=+0$, and takes minimum and maximum there~\cite{Ando2020-om}.}
{$\Delta j_c(h)$ near $T_c$
seems to be at variance with the intrinsic SDE~\cite{Ando2020-om}.}
This point might be overcome by considering the effect of vortices, which is left as an intriguing future issue.

% \AD{Note that there are several branches for $j(q)$.
% When experimentally increasing the applied electric current from $j=0$, there are two possibilities:
% When $j$ reaches the critical value for the local minima, (i) there occurs a first-order phase transition into the other branch~\cite{Samkhin2017@@}; or (ii) superconductivity is destroyed and metal-superconductor transition takes place.
% Both cases might occur depending on the experimental setups.
% The values of $\Delta j_c$ and $r$ discussed in the main text are for the scenario (i).
% On the other hand, even for the case (ii), the main conclusions do not change.
% Indeed, for the scenario (ii), SDE occurs due to the asymmetry of the most stable branch.
% Accoding to Fig.(b), $\Delta j_c^{(ii)}>0$ with a large asymmetry with $r^{(ii)}\sim 1$ is realized, while $\Delta j_c^{(ii)}<0$ with $r^{(ii)}\sim -1$ is realized in Fig.(c).
% Thus, a large SDE and the sign reversals around the crossover line are also obtained in this case.
% }

\begin{acknowledgments}
We appreciate helpful discussions with T. Ono, Y. Miyasaka, R. Kawarazaki, H. Narita, and H. Watanabe.
This work was supported by JSPS KAKENHI (Grants No. JP18H05227, No. JP18H01178, No. 20H05159, No. 21K13880, and No. 21J14804), JSPS research fellowship, WISE Program MEXT, and SPIRITS 2020 of Kyoto University.

\textit{Note added.}---
During finalizing the manuscript, we became aware of independent overlapping works.
A recent arXiv post by N. Yuan and L. Fu~\cite{Yuan2021-pz} studies the depairing currrent of the Rashba-Zeeman model mainly using the GL theory.
However, sign reversals of the SDE are not obtained.
{The work by J. He and N. Nagaosa {\it et al.}~\cite{He2021} studies the related topic independently of ours.}
We thank J. He for coordinating submission to arXiv.
\end{acknowledgments}

%\bibliography{ref_for_addition,ref_pp}

%%%%%%%%%%%%%%%%%%%%%%%%%%%%%%%%%%%%%%%%5
% \section{Should be Added}

% \AD{It is known that the high-field helical superconductivity is fragile against impurities, while the low-field state is robust against moderate disorders~\cite{Michaeli2012-gl,Dimitrova2007-hp,Houzet2015-iy,Samokhin2008-nv,Smidman2017-hb}.
% Thus, the sign reversal of $\Delta j_c$ accompanied by the crossover is not expected in dirty-limit superconductors.
% On the other hand, a large SDE as well as the sign reversals caused by the nonreciprocal Landau critical momenta may survive. (60)}

\section{Formulation to evaluate the electric current}
Here, we show the details of the formulation to calculate the current expectation values in the superconducting states.
The free energy per unit volume in the superconducting state is given by
\begin{align}
    \Omega(\Delta,q)&\equiv-\frac{T}{V}\ln\Tr e^{-\hat{H}_{\text{MF}}^q(\Delta)/T}\notag\\
    &=\frac{1}{2V}\sum_{\bm{k}}\tr_N\left[\frac{\Delta^2}{U}+H_N(\bm{k})\right]\notag\\
    &\qquad-\frac{T}{2V}\sum_{\bm{k}}\tr\left[\ln(1+e^{-H(\bm{k},q)/T})\right].
    \label{eq:Omega_def}
\end{align}
Here, $\tr_N$ represents the trace over the spin degrees of freedom, while $\tr$ represents that over both the spin and the Nambu degrees of freedom.
$V=L_xL_y$ represents the system size with $L_i$ the diameter in the   $i=x,y$ direction.
We introduced the Bogoliubov-de Genens (BdG) Hamiltonian $H(\bm{k},q)$ by
\begin{gather}
    \hat{H}_{\text{MF}}^q(\Delta)=\frac{1}{2}\sum_{\bm{k}}\Psi^\dagger(\bm{k},q)H(\bm{k},q)\Psi(\bm{k},q)+\text{const.},\label{eq:temp1}\\
    H(\bm{k},q)=\begin{pmatrix}H_N(\bm{k}+\bm{q})&\Delta i\sigma_y\\-\Delta i\sigma_y &-H_N^T(-\bm{k})\end{pmatrix},\label{eq:BdG_Hamiltonian}
\end{gather}
with the momentum $\bm{q}=q\hat{x}$ and the Nambu spinor $\Psi(\bm{k},q)^\dagger=(c^\dagger_{\bm{k}+\bm{q}\uparrow},c^\dagger_{\bm{k}+\bm{q}\downarrow},c_{-\bm{k}\uparrow},c_{-\bm{k}\downarrow})$.
Here, we choose $q$ to be compatible with the periodic boundary conditions, $q\in2\pi\mathbb{Z}/L_x$.
The constant term in Eq.~\eqref{eq:temp1} is equivalent to the first term of Eq.~\eqref{eq:Omega_def}.
The normal-state Bloch Hamiltonian is given by $H_N(\bm{k})=\xi(\bm{k})+(\bm{g}(\bm{k})-\bm{h})\cdot\bm{\sigma}$.

The electric current (the sheet current density) is defined by
\begin{gather}
j(\Delta,q)=\frac{\tr[\hat{j}_x e^{-\hat{H}^q_{\text{MF}}(\Delta)/T}]}{\tr[e^{-\hat{H}^q_{\text{MF}}(\Delta)/T}]}.
\end{gather}
Here, the current operator is given by
\begin{align}     \hat{j}_x&=\frac{1}{V}\sum_{\bm{k}\sigma\sigma'}\partial_{k_x}H_N(\bm{k})_{\sigma\sigma'}c^\dagger_{\bm{k}\sigma}c_{\bm{k}\sigma'}\\
       &=\frac{1}{V}\sum_{\bm{k}}\Psi^\dagger(\bm{k},q)\partial_qH(\bm{k},q)\Psi(\bm{k},q).
\end{align} 
After some calculations, we obtain
\begin{align}
    j(\Delta,q)&=\frac{1}{V}\sum_{\bm{k}}\tr[\partial_qH(\bm{k},q)f(H(\bm{k},q))]\\
    &=2\partial_q\Omega(\Delta,q),\label{eq:current_temp}
\end{align}
with the Fermi distribution function $f(\epsilon)=(e^{\epsilon/T}+1)^{-1}$.

The gap equation is given by
\begin{equation}
    \partial_{\Delta}\Omega(\Delta,q)=0,\label{eq:gapeq_free}
\end{equation}
which determines the pair potential $\Delta$ self-consistently.
The solution is written as $\Delta(q)$, and satisfies Eq.~\eqref{eq:gapeq_free}, or equivalently,
\begin{equation}
\Delta(q)=-\frac{U}{V}\sum_{k}\braket{c_{-k\downarrow}c_{k+q\uparrow}}|_{\Delta=\Delta(q)}.
\end{equation}
By using $\Delta(q)$, the condensation energy defined in the main text is written as
\begin{equation}
    F(q)=\Omega(\Delta(q),q)-\Omega(0,q),
\end{equation}
where $\Omega(0,q)=\Omega(0,0)$ holds as is easily confirmed with Eqs.~\eqref{eq:Omega_def} and \eqref{eq:BdG_Hamiltonian}.
By using Eqs.~\eqref{eq:current_temp} and \eqref{eq:gapeq_free}, we obtain 
\begin{align}
2\partial_qF(q)
&=\lim_{\Delta\to\Delta(q)}\bigl[2\partial_{q}\Omega(\Delta,q)+2\partial_q\Delta(q)\,\partial_{\Delta}\Omega(\Delta,q)\bigr]\notag\\
&=j(\Delta(q),q)\notag\\
&\equiv j(q).
\end{align}
The obtained equality goes along with the standard expression $j=-\partial_A\Omega$ with $A$ the uniform vector potential since $q$ changes by $-2\delta A$ when $A$ changes by $\delta A$.

\section{GL analysis}
Here we show the details of the GL analysis of the superconducting diode effect.
Let us start from the expression
\begin{equation}
    f(\Delta,q)=\alpha(q)\Delta^2+\frac{\beta(q)}{2}\Delta^4,
    \label{eq:GL_SM}
\end{equation}
keeping the order parameter of the form $\Delta(x)\propto e^{iqx}$ in mind.
The coefficients are given by
\begin{gather}
    \alpha(q)=-\talpha_0+\frac{\talpha_2}{2}(q-q_0)^2+\frac{\alpha_3}{6}(q-q_0)^3\notag,\\
    \beta(q)=\tbeta_0+\beta_1(q-q_0).\label{eq:GL_coeff_SM}
\end{gather}
Here we omit the tilde of $\tilde{q}_0$ in the main text for simplicity, and redefine $\talpha_0\to-\talpha_0$.
The order parameter is optimized by
\begin{equation}
     \partial f/\partial{\Delta^2}=\alpha(q)+\beta(q)\Delta^2=0.
\end{equation}
Assuming $\beta(q)>0$ for the range of $q$ we are interested in, $\Delta$ has a nontrivial real solution only when $\alpha(q)<0$.
Thus, the GL free energy is given by
\begin{equation}
    f(q)=f(\Delta(q),q)=-\frac{\alpha(q)^2}{2\beta(q)}\theta(-\alpha(q)),
\end{equation}
with $\theta(x)$ the Heaviside step function.
Since the minimum of $\alpha(q)$ is $-\talpha_0$, the  transition from the normal to helical superconducting state occurs when the sign of $\talpha_0$ changes from negative to positive as lowering the temperature.
Thus, we conclude $\talpha_0\propto T_c-T$.

We first consider the case $\alpha_3=\beta_1=0$.
The supercurrent is given by $j(q)=2\partial_qf(q)$, 
\begin{equation}
    \tbeta_0j(q)/2=-\partial_q\alpha(q)^2/2=-\alpha(q)\partial_q\alpha(q).
\end{equation}
This is an odd function of $q-q_0$, and thus the critical current is reciprocal.
Actually, The maximum and minimum of $j(q)$ are achieved at $q=q_c$ satisfying
\begin{align}
    0&=[\partial_{q_c}\tbeta_0j(q_c)/2]_{\alpha_3=\beta_1=0}\notag\\
&=\frac{3\talpha_2^2}{2}\left(\frac{2\talpha_0}{3\talpha_2}-(q_c-q_0)^2\right).
\end{align}
Accordingly, $|q_c-q_0|$ scales as $\sqrt{T_c-T}$, as is the inverse of the coherence length.
Thus, the reciprocal critical current is given by
\begin{align}
    [j_{c+}]_{\alpha_3=\beta_1=0}&=[-j_{c-}]_{\alpha_3=\beta_1=0}\notag\\
    &=\frac{4\sqrt{6\talpha_2}}{9\tbeta_0}\talpha_0^{3/2}.
\end{align}
Note that this coincides with $\bar{j}_c$ up to first order in $\alpha_3$ and $\beta_1$.
Thus, the well-known scaling law $j_c\sim (T_c-T)^{3/2}$ is reproduced for $\bar{j}_c$.

Let us consider the first-order change caused by $\alpha_3$ and $\beta_1$.
We obtain
\begin{align}
    &\partial_{\alpha_3}[\tbeta_0 j(q)/2]\Bigr|_{\alpha_3=\beta_1=0}\notag\\
    &=\frac{5\talpha_2(q-q_0)^2}{12}\left(\frac{6\talpha_0}{5\talpha_2}-(q-q_0)^2\right),
\end{align}
and
\begin{align}
    &\partial_{\beta_1}[\tbeta_0j(q)/2]\Bigr|_{\alpha_3=\beta_1=0}\notag\\
    &=\frac{5\talpha_2^2}{8\tbeta_0}\left((q-q_0)^2-\frac{2\talpha_0}{\talpha_2}\right)\left((q-q_0)^2-\frac{2\talpha_0}{5\talpha_2}\right).
\end{align}
When the critical current $j_{c+}$ is realized at $q_{c+}=q_c+\delta q_c$, we obtain up to first order in $\alpha_3$ and $\beta_1$,
\begin{align}
    j(q_{c+})&=[j+\delta j](q_c+\delta q_c)\notag\\
    &=j_{c0}+\alpha_3[\partial_{\alpha_3}j(q_c)]_{\alpha_3=\beta_1=0}\notag\\
    &\quad+\beta_1[\partial_{\beta_1}j(q_c)]_{\alpha_3=\beta_1=0}
    \notag\\
    &\quad+[\partial_{q_c}j(q_c)]_{\alpha_3=\beta_1=0}\delta q_c\notag\\
    &=j_{c0}+\alpha_3[\partial_{\alpha_3}j(q_c)]_{\alpha_3=\beta_1=0}\notag\\
    &\quad+\beta_1[\partial_{\beta_1}j(q_c)]_{\alpha_3=\beta_1=0}.
\end{align}
Here, $j_{c0}$ represents $[j(q_c)]_{\alpha_3=\beta_1=0}$.
Thus, we obtain the nonreciprocal component of the critical current,
\begin{align}
    \Delta j_c
    &=2\alpha_3[\partial_{\alpha_3}j(q_c)]_{\alpha_3=\beta_1=0}
+2\beta_1[\partial_{\beta_1}j(q_c)]_{\alpha_3=\beta_1=0}
    \notag\\
    &=\frac{16\talpha_0^2}{27\tbeta_0\talpha_2}\alpha_3-\frac{8\talpha_0^2}{9\tbeta_0^2}\beta_1.
\end{align}
This scales as $\Delta j_c\sim (T_c-T)^2$.

\begin{figure*}[t]
    \centering
    \begin{tabular}{llll}
    (a)&(b)&(c)&(d)\\
       \includegraphics[width=0.25\textwidth]{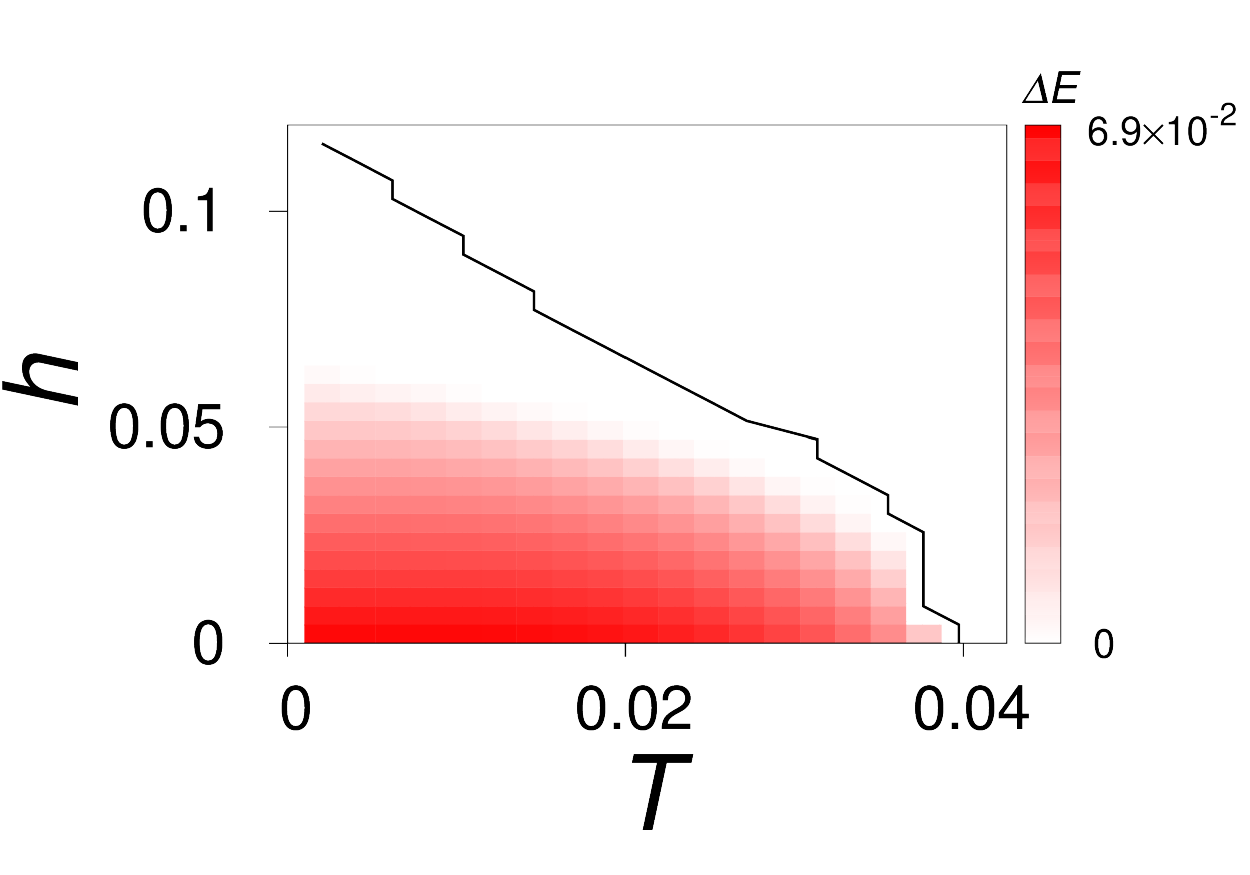} &
     \includegraphics[width=0.25\textwidth]{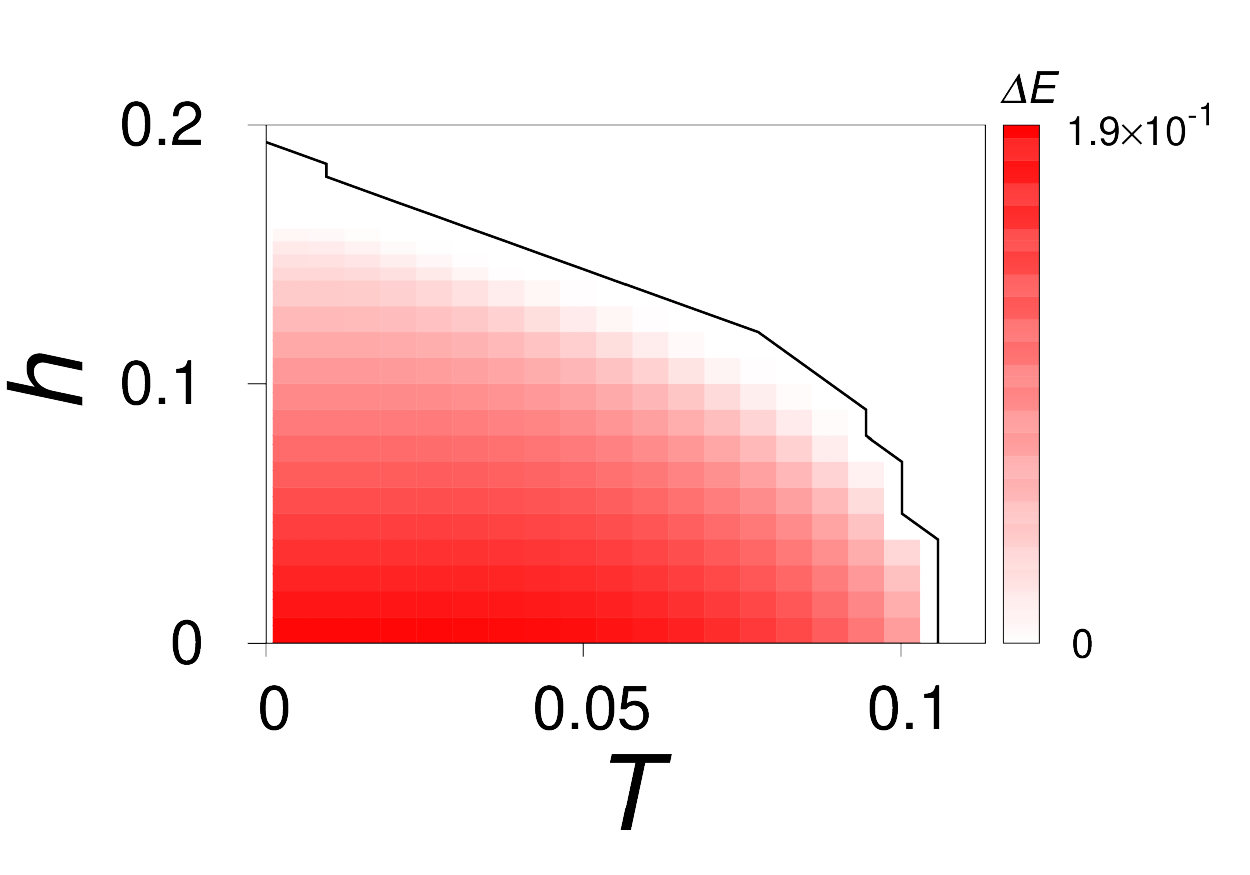} &
     \includegraphics[width=0.25\textwidth]{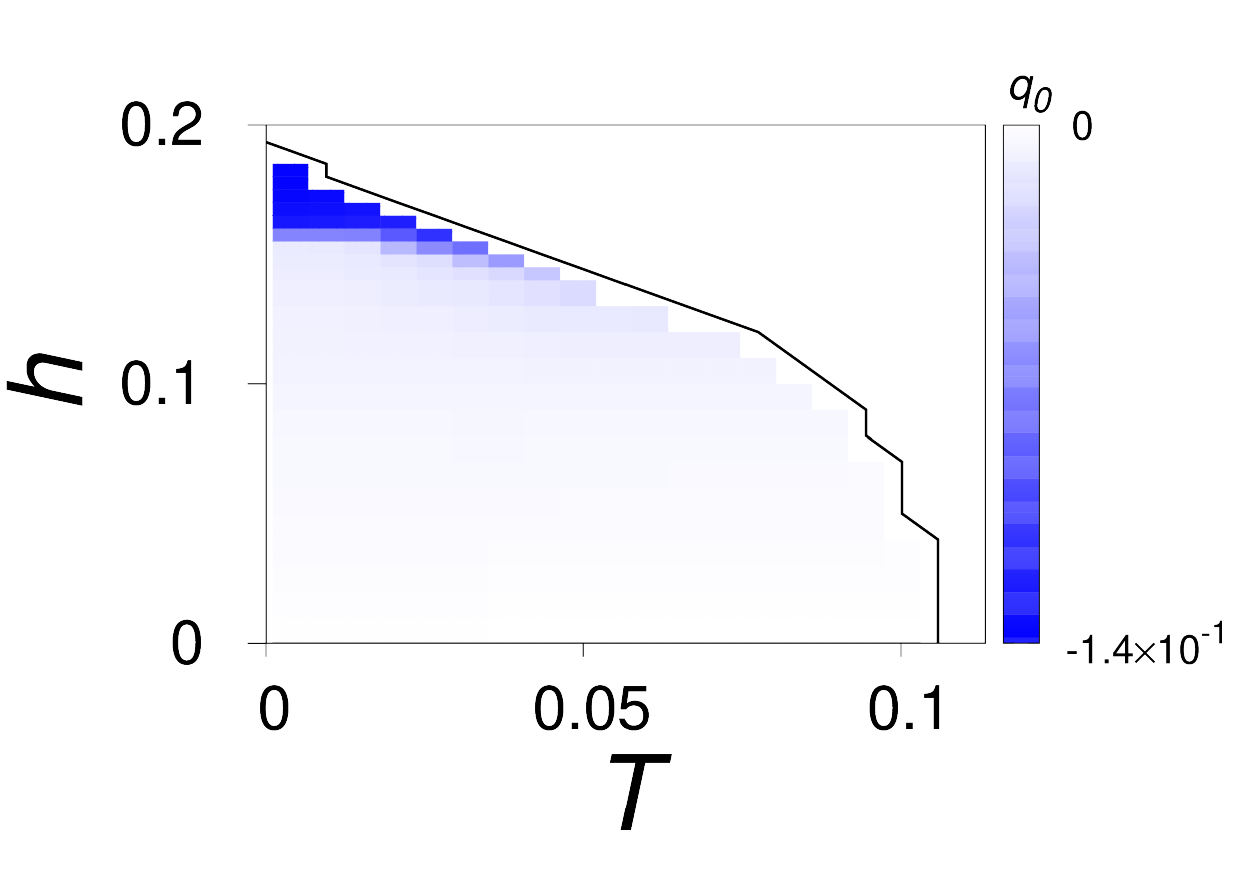}&
    \includegraphics[width=0.25\textwidth]{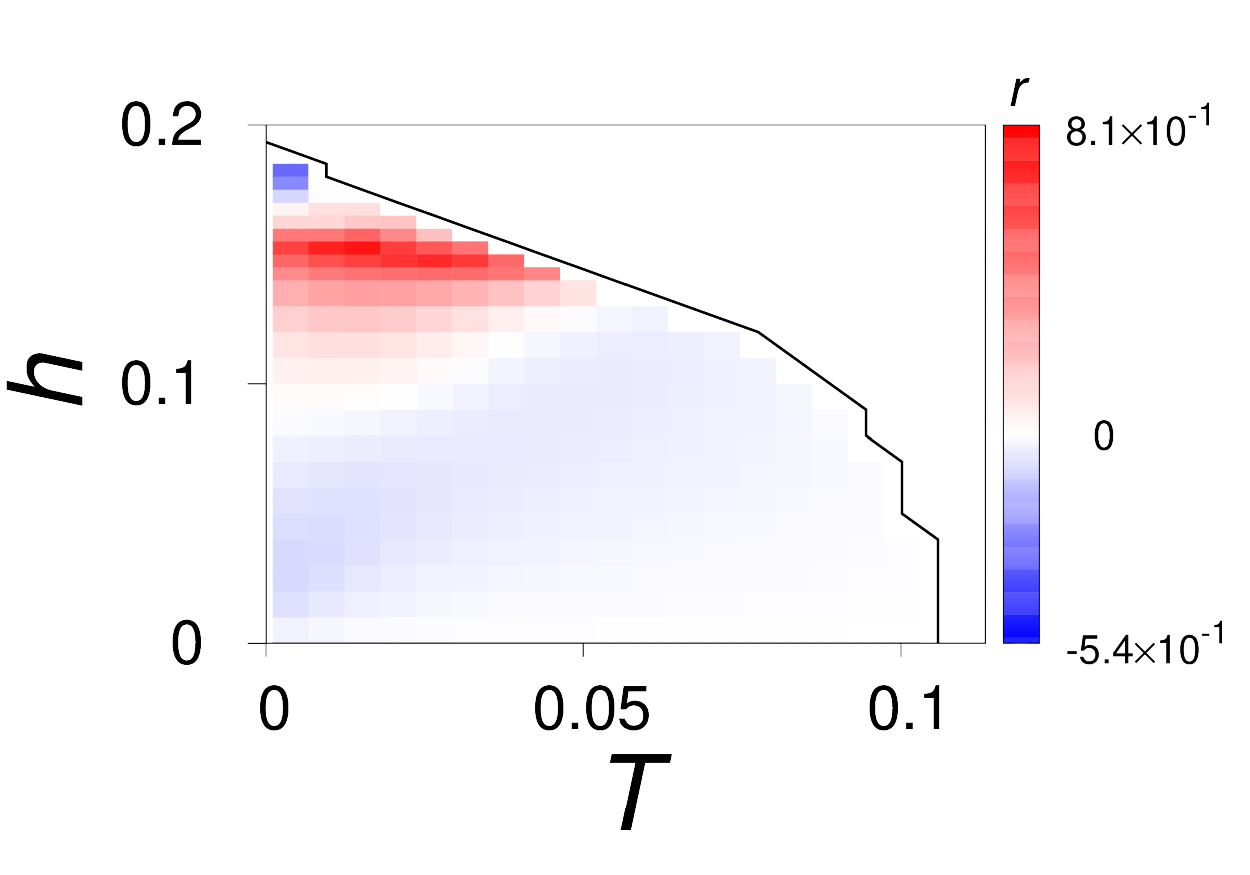}
     \end{tabular}
    \caption{Temperature and magnetic-field dependence of (a),(b) $\Delta E$, (c) $q_0$, and (d) $r$.
    $t_2=0$ for the panel (a), while $t_2=0.2$ for the panels (b)-(d).}
    \label{fig:figures_SM}
\end{figure*}

\section{Calculation details and Phase diagrams}
Here we explain the details of the numerical calculations and show some additional figures related to the phase diagrams.
All the calculations for the figures in the main text and those presented here are done with $L_x=6000$ and $L_y=200$.
Exceptionally, we adopt $L_x=12000$ for Fig.~5 and $T>0.03$ in Fig.~2 to reduce the finite-size effect.
To obtain $j_{c\pm}$, $j(q)$ is maximized/minimized among $q\in 2\pi\mathbb{Z}/L_x$.
The normalization of Figs.~5~(a)-(d) is done with $\max[j_{c+},|j_{c-}|]$, $|\min_qF(q)|$, and $\max[0.003,\max_q\Delta E(q)]$ for $j(q)$, $F(q)$, and $\Delta E(q)$, respectively.
Note that we show in Figs.~5~(a)-(d) only the most stable state that minimizes $F(q)$ for each $q$.
In particular, there is a metastable superconducting state for smaller (larger) $q$'s of the left (right) peaks in Fig.~5~(d).
The supercurrent sustained by these states might be observed when the experimental time scale is small.
For Figs.~5 (a)-(d), it is confirmed that the superconducting solution is (if any) unique for each $q$.

In Figs.~\ref{fig:figures_SM}~(a) and (b), we show the temperature and magnetic-filed dependence of the minimum excitation energy $\Delta E(q_0)$ for $t_2=0$ and $0.2$, respectively.
The spectrum becomes gapless in the high-field helical superconducting states, which can be detected by scanning tunneling microscopy.
Figures~\ref{fig:figures_SM}~(c) and (d) show $q_0$ and $r$ for $t_2=0.2$.
In Figs.~\ref{fig:figures_SM}~(c) and (d), the crossover line is seen to be $h\sim 0.17$, and a huge nonreciprocal nature $r\sim 0.8$ is observed.
To be precise, $\Delta j_c$ becomes positive in a tiny region near $h\sim0.19$ and $T\sim0$ for $t_2=0.2$. However, this is probably due to the peculiarity of the model, where the Lifshitz transition of the outer Fermi surface occurs around $h\sim0.185$.

\section{Evolution of Landau critical momenta}
\begin{figure}[h]
    \centering
   \includegraphics[width=0.45\textwidth]{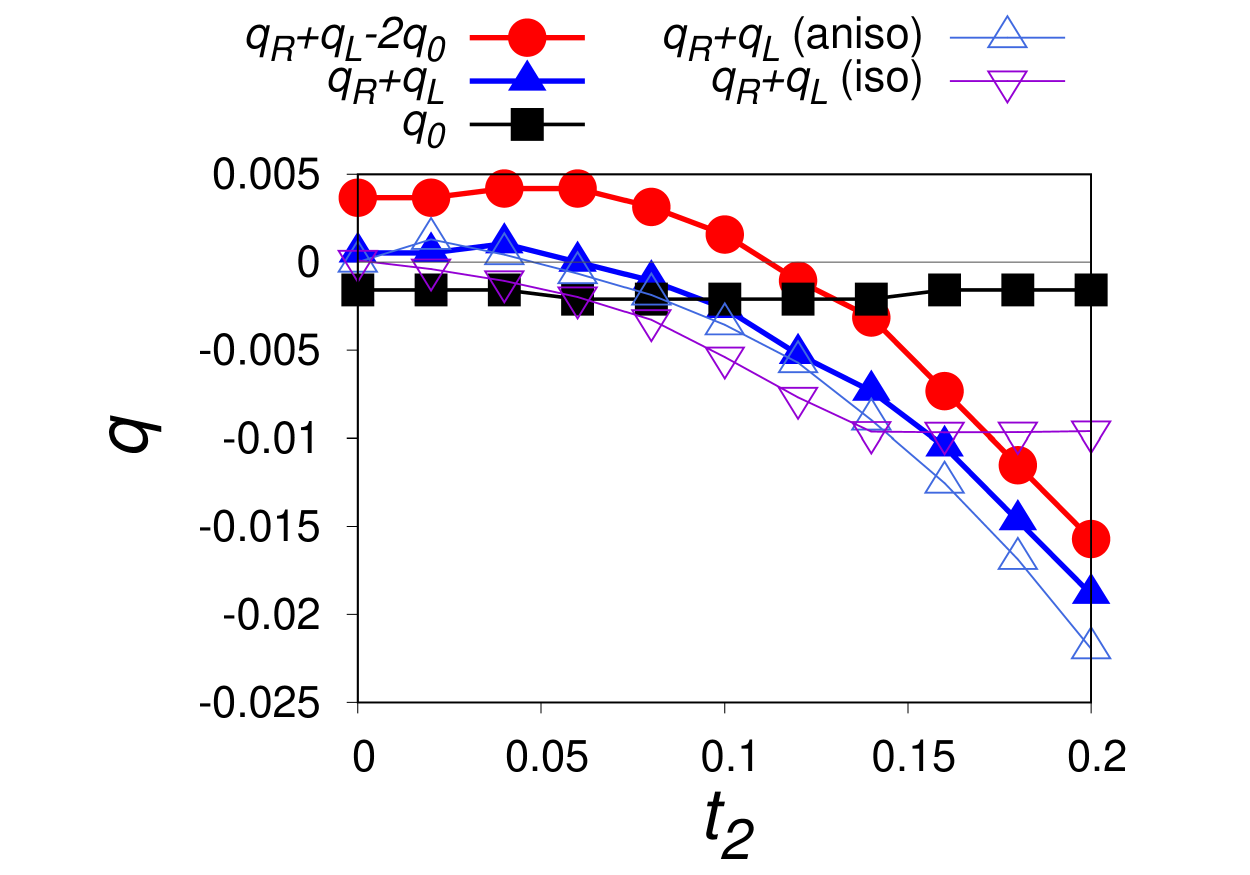}
  \caption{$t_2$ dependence of $q_R+q_L$, $q_0$, and their combinations. The red closed circles, blue closed triangles, black closed squares indicate $q_R+q_L-2q_0$, $q_R+q_L$, $q_0$ evaluated from $j(q)$, respectively. The open sky-blue triangles and open purple inverted triangles indicate $q_R+q_L$ calculated from Eq.~\eqref{eq:Landau_CM} and that with the isotropic simplification $\bm{q}_\chi(\bm{k})\to q_\chi\hat{x}$, respectively.
    }
    \label{fig:qR_vs_t2}
\end{figure}

As discussed in the main text, the sign reversal of $\Delta j_c$ by $t_2$ can be understood based on the nonreciprocity of the Landau critical momenta.
In Fig.~\ref{fig:qR_vs_t2}, we show $q_R+q_L$ and $q_0$ obtained from $j(q)$ at $h=0.03$ and $T=0.001$, with varying 
$t_2$ from $0$ to $0.2$.
The sign reversal of $q_R+q_L-2q_0$ (red closed circles) naturally explains that of $\Delta j_c$.
For large values of $t_2$, $q_R+q_L-2q_0$ is dominated by the nonreciprocal Landau critical momentum $q_R+q_L$ (blue closed triangles), while it is dominated by $q_0$ (black closed squares) for small values of $t_2$.
It should be noted that a large SDE is obtained for $t_2\sim 0.2$, although $q_R+q_L$ and $-2q_0$ contribute destructively.

To understand the behavior of $q_R+q_L$, we discuss
the Landau critical momenta with the help of the single-band formula $|\bm{q}\cdot\bm{v}/2|=\Delta$.
In our case, $\bm{q}$ should be replaced with $\bm{q}-\bm{q}_\chi(\bm{k})$ for each band, and we obtain,
\begin{align}
    \Delta&=\max_{\chi=\pm,\ \bm{k}=\bm{k}_F^\chi}\left|\frac{q\,\hat{x}-\bm{q}_\chi(\bm{k})}{2}\cdot\bm{v}_\chi(\bm{k})\right|\notag\\
    &=\max_{\chi=\pm,\ \bm{k}=\bm{k}_F^\chi}\left|qv_{\chi,x}(\bm{k})/2-\chi h\hat{g}_y(\bm{k})\right|,\label{eq:Landau_CM}
\end{align}
whose positive and negative solutions for $q$ correspond to $q_R$ and $q_L$, respectively.
Here, $\bm{k}_F^\chi$ specifies the $\bm{k}$ points on the Fermi surface with the helicity $\chi$, while $\hat{g}(\bm{k})$ is the unit vector parallel to $\bm{g}(\bm{k})$.
Equation~\eqref{eq:Landau_CM} well reproduces the result for $q_R+q_L$, as shown by skyblue open triangles in Fig.~\ref{fig:qR_vs_t2}.
To go further, let us simplify the expression by replacing $\bm{q}_\chi(\bm{k})$ in the first line of Eq.~\eqref{eq:Landau_CM} with its average on the Fermi surface: $q_\chi\hat{x}\equiv \braket{\bm{q}_{\chi}(\bm{k}_{F,\chi})}$.
We obtain for $h>0$ [see the next section for the derivation],
\begin{align}
    q_R+q_L&=\begin{cases}\begin{array}{l}
        q_++q_-\\
         \ +2\left(\frac{\Delta}{v_-}-\frac{\Delta}{v_+}\right),
    \end{array}&(|\delta v|/\bar{v}\lesssim h/\Delta)\\
 2q_+,&(\delta v/\bar{v}\gtrsim h/\Delta)\\
 2q_-,&(\delta v/\bar{v}\lesssim -h/\Delta)\end{cases}
 \label{eq:Landau_iso}
\end{align}
whose helicities are interchanged for $h<0$.
Here, we defined $v_\chi\equiv \max_{\bm{k}_{F}^{\chi}}v_{\chi,x}(\bm{k}_{F}^{\chi})$, $\delta v=v_+-v_-$ and $\bar{v}\sim (v_++v_-)/2$.
Equation~\eqref{eq:Landau_iso} qualitatively agrees with $q_R+q_L$ for $t_2\lesssim 0.15$, as shown by the open purple inverted triangles in Fig.~\ref{fig:qR_vs_t2}.
In this regime, we have small $\delta v$ and the first line of Eq.~\eqref{eq:Landau_iso} is applied.
Since $q_\pm\sim \pm 2h/v_\pm$, the difference of the Fermi velocities $\delta v$ plays a key role to obtain a large $q_R+q_L$.
It is expected that the anisotropy of the system is advantageous to obtain a large value of $\delta v$.
On the other hand, Eq.~\eqref{eq:Landau_iso} underestimates $q_R+q_L$ around $t_2=0.2$, where the third line of Eq.~\eqref{eq:Landau_iso} is applied.
This indicates that the isotropic simplification $\bm{q}_\chi(\bm{k})\to q_\chi$ is not valid for strongly anisotropic systems with large $t_2$.
Thus, overall, large anisotropy of the system is expected to be the key to obtain a large SDE.

\begin{figure}[h]
    \centering
\includegraphics[width=0.45\textwidth]{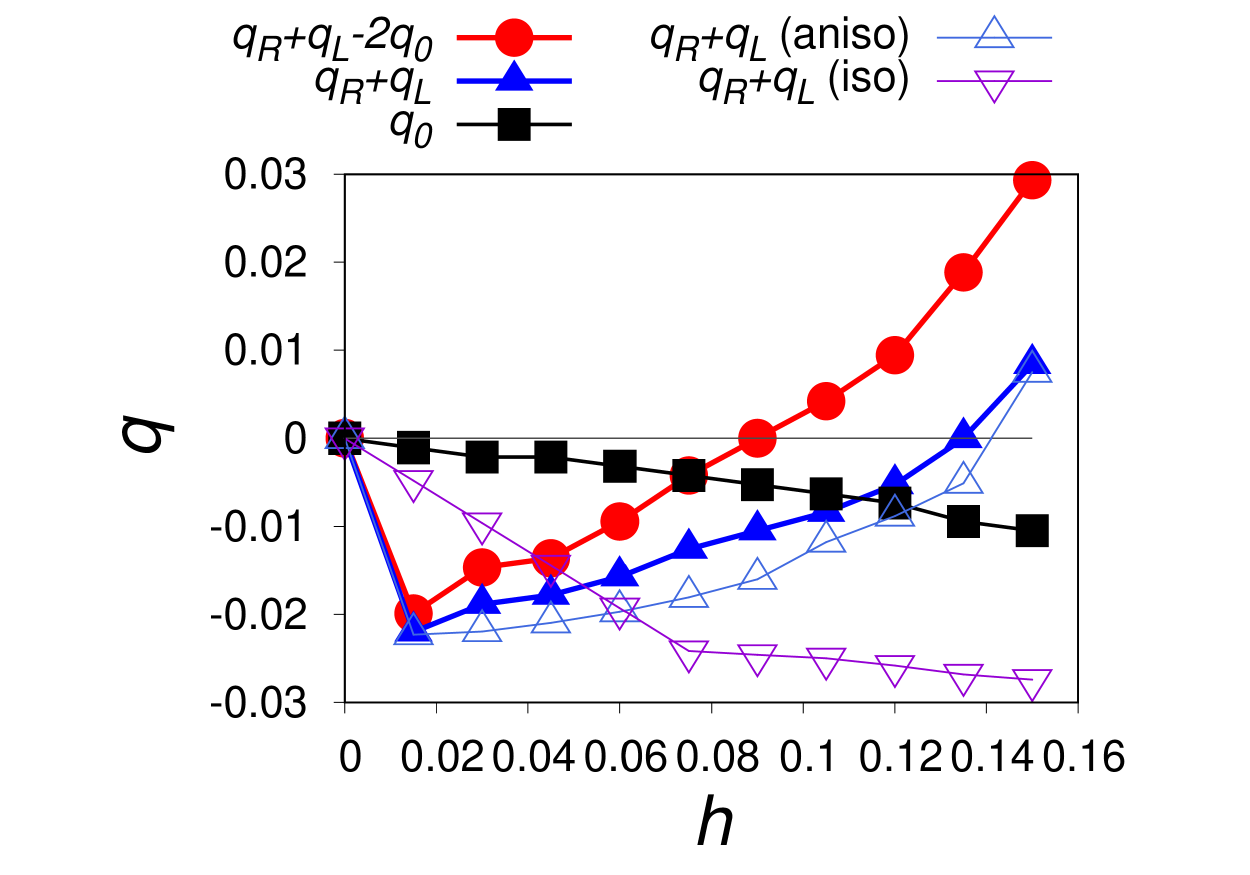}
    \caption{Magnetic field dependence of $q_R+q_L$, $q_0$ and their combinations for $t_2=0.2$ and $T=0.001$.
    The notations are the same as Fig.~\ref{fig:qR_vs_t2}.
    }
    \label{fig:qR_vs_h}
\end{figure}
In Fig.~\ref{fig:qR_vs_h}, we show the magnetic-field dependence of $q_R+q_L$, $q_0$ and  their combinations for $t_2=0.2$.
The notations are the same as those of Fig.~\ref{fig:qR_vs_t2}.
While $q_R+q_L$ is nonmonotonic, $-2q_0$ grows linearly and finally the sign reversal of $q_R+q_L-2q_0$ occurs.
``$q_R+q_L$ (aniso)", i.e. Eq.~\eqref{eq:Landau_CM}, qualitatively captures the behavior of $q_R+q_L$, whose slight deviation is probably due to the higher-order corrections of $h$.
The isotropic simplification does not work for $t_2=0.2$ as is clear in Fig.~\ref{fig:qR_vs_h}.
The sign reversal of $q_R+q_L-2q_0$ is the origin of that of $\Delta j_c$ in the phase diagram for $t_2=0.2$ under moderate magnetic fields.

\subsection{Derivation of Eq.~\eqref{eq:Landau_iso}}
\label{sec:Landau_iso_derivation}
Here, we derive Eq.~\eqref{eq:Landau_iso}.
By using $\bm{q}_\chi(\bm{k})\to q_\chi\hat{x}$, we obtain
\begin{align}
    2\Delta&=\max_{\chi=\pm}\Bigl[|q-q_\chi|\max_{\bm{k}_{F}^\chi}|v_{\chi,x}(\bm{k})|\Bigr]\\
    &=\max_{\chi=\pm}\Bigl[|q-q_\chi|v_\chi\Bigr].
\end{align}
Let us first consider the positive solution $q=q_R>0$.
We also fix $h>0$.
Then, we obtain
\begin{align}
    2\Delta&=\max\Bigl[(q_R+|q_-|)v_-,\,(q_R-|q_+|)v_+\Bigr].
\end{align}
When $(q_R+|q_-|)v_->(q_R-|q_+|)v_+$, we obtain
\begin{equation}
    q_R=q_-+\frac{2\Delta}{v_-}.
\end{equation}
The consistency can be checked as follows.
The above inequality reads
\begin{align}
\delta v q_R<|q_-|v_-+|q_+|v_+\sim 2h.
\end{align}
Thus, this solution is valid for
\begin{equation}
    2h\gtrsim \delta v\frac{2(\Delta-h)}{v_-}\sim\frac{2\delta v(\Delta-h)}{\bar{v}}.
\end{equation}
Considering only the linear dependence in $h$, we obtain
\begin{equation}
q_R=q_-+\frac{2\Delta}{v_-},\quad \frac{\delta v}{\bar{v}}\lesssim\frac{h}{\Delta}.    
\end{equation}
In the same way, we obtain
\begin{equation}
    q_R=q_++\frac{2\Delta}{v_+},
\end{equation}
for $\delta v q_R\gtrsim 2h$, i.e. $\delta v/\bar{v}\gtrsim h/\Delta$.
The negative solutions $q=q_L<0$ are obtained as follows:
\begin{equation}
    q_L=\begin{cases}
    q_+-\frac{2\Delta}{v_+},&\frac{\delta v}{\bar{v}}\gtrsim\frac{-h}{\Delta}\\
    q_--\frac{2\Delta}{v_-}.&\frac{\delta v}{\bar{v}}\lesssim\frac{-h}{\Delta}
    \end{cases}
\end{equation}
Summing up $q_R$ and $q_L$, we obtain Eq.~\eqref{eq:Landau_iso}.

\end{document}